\documentclass[aps,prb,twocolumn,
	amsmath,amssymb,amsfonts,floatfix,
	longbibliography,superscriptaddress,nobibnotes]{revtex4-2}

\usepackage[utf8]{inputenc}
\usepackage[T1]{fontenc}
\usepackage{graphicx} 
\usepackage{bm} 
\usepackage{ulem}

\usepackage[hypertexnames=false,colorlinks=true,citecolor=blue,
			linkcolor=blue,urlcolor=blue]{hyperref}

\usepackage{xfrac}

\usepackage{xcolor}
\definecolor{mygreen}{rgb}{0.0, 0.6, 0.0}

\newcommand{\evec}{\text{e}}

\graphicspath{{fig/}{./fig/}{.}}

\begin{document}

\title{Chiral phonon in the cubic system based on the Laves phase of $A$Bi$_{2}$ ($A=$K, Rb, Cs)}

\author{Surajit Basak}
\email[e-mail: ]{surajit.basak@ifj.edu.pl}
\affiliation{\mbox{Institute of Nuclear Physics, Polish Academy of Sciences, W. E. Radzikowskiego 152, PL-31342 Krak\'{o}w, Poland}}

\author{Przemys\l{}aw Piekarz}
\affiliation{\mbox{Institute of Nuclear Physics, Polish Academy of Sciences, W. E. Radzikowskiego 152, PL-31342 Krak\'{o}w, Poland}}

\author{Andrzej Ptok}
\email[e-mail: ]{aptok@mmj.pl}
\affiliation{\mbox{Institute of Nuclear Physics, Polish Academy of Sciences, W. E. Radzikowskiego 152, PL-31342 Krak\'{o}w, Poland}}

\date{\today}

\begin{abstract}
$A$B$_{2}$ ($A=$K, Rb, Cs) compounds crystallize in the cubic Laves phase (symmetry Fd$\bar{3}$m).
The geometry of the crystal structure allows the realization of chiral phonons, which are associated with the circulation of atoms around their equilibrium positions. 
Due to the inversion symmetry and time reversal symmetry, total pseudo-angular momentum (PAM) of the system vanishes.
We show that the doping of these system can lead to a new phase with symmetry F$\bar{4}3$m.
New systems (KRbBi$_{4}$ and RbCsBi$_{4}$) do not exhibit soft modes (are stable dynamically).
Due to the inversion symmetry breaking, realized chiral phonon modes posses a non-zero total PAM.
In both type of systems the chiral phonons are realized for the wavectors at the edge of the Brillouine zone.
This study explores the possibility of chiral phonon engineering via doping, and predicts two new materials.
Discussing the problem opens a new way to study the phonon Hall effect.
\end{abstract}

\maketitle

\paragraph*{Introduction.---}
Chirality, a property, which distinguishes a system from its mirror image, plays an important role in physics.
In lattice dynamics, chirality can be discuss in context of the chiral phonons, i.e., vibration modes associated with the circular motion of the atoms around the equilibrium position~\cite{zhang.niu.15}.
Chiral phonons were studied in many two dimensional (2D) lattices, e.g., honeycomb lattice~\cite{zhang.niu.15,liu.lian.17}, kagome lattice~\cite{chen.wu.19}, or moir\'{e} superlattices~\cite{suri.wang.21}.
Recently, chiral phonons were also reported in many three dimensional (3D) materials, e.g.:
transition metal \mbox{dichalcogenides~\cite{chen.zheng.15,zhu.yi.18,liu.vanbaren.19,chen.lu.19}} and their heterostructures~\cite{delhomme.vaclavkova.20,zhang.srivastava.20,maity.mostofi.22}, 
pervoskites~\cite{nova.cartella.17,juraschek.fechner.17,juraschek.spaldin.19,li.fauque.20}, 
graphene/hexagonal boron nitride heterostructure~\cite{gao.zhang.18},
2D magnets (CrBr$_{3}$~\cite{yin.ulman.21} or Fe$_{3}$GeTe$_{2}$~\cite{du.tang.19}),
cuprates~\cite{grissonnanche.theriault.20},
CoSn-like systems~\cite{ptok.kobialka.21}, 
ternary YAlSi compound~\cite{basak.ptok.22},
chiral systems ($A$Bi-like compounds~\cite{skorka.kapcia.22}, $\alpha$-HgS~\cite{ishito.mao.21} or SiO$_{4}$~\cite{wang.li.21}),
and magnetic topological insulators $T$Bi$_{2}$Te$_{4}$~\cite{kobialka.sternik.22}.
Due to their extraordinary properties (e.g., realization of the phonon Hall effect~\cite{stohm.rikken.05,zhang.ren.11,sugii.shimozawa.17,kasahara.sugii.18,zhang.zhang.19,boulanger.grissonnanche.20,yokoi.ma.21,zhang.xu.21}), 
it has attracted a lot of theoretical and experimental attentions.

\begin{figure}[!b]
\includegraphics[width=\linewidth]{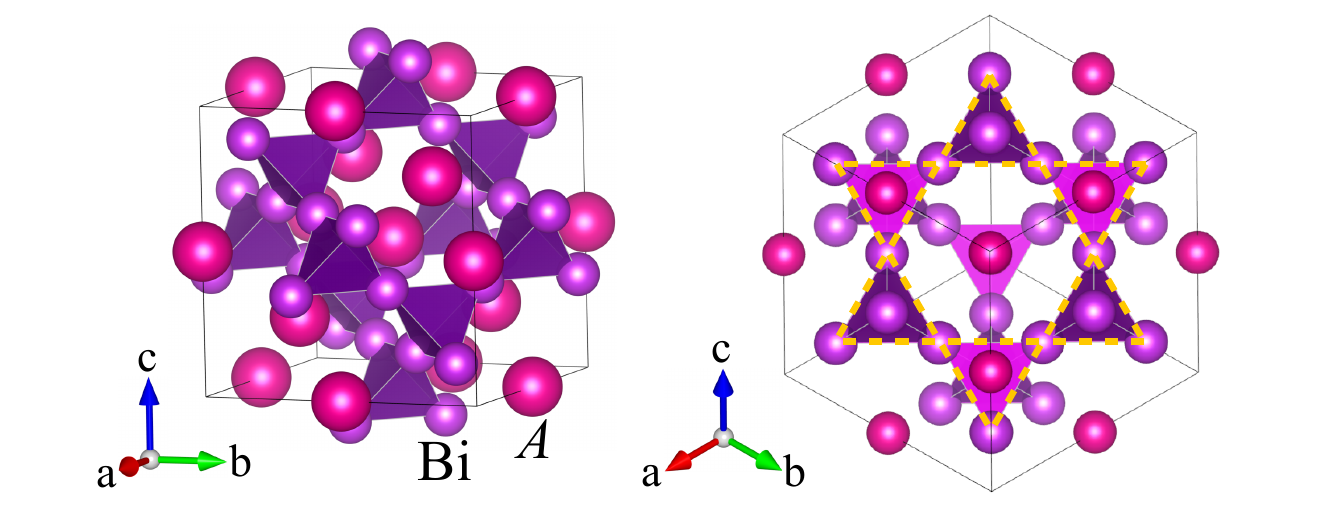}
\caption{
(a) Conventional cell of $A$Bi$_{2}$ in cubic Leves phase.
(b) Crystall structure along (111) direction presenting kagome net (dashed red line) of corner-shared tetrahedra of Bi atoms.
}
\label{fig.crys}
\end{figure}

\begin{figure*}[!t]
\includegraphics[width=\linewidth]{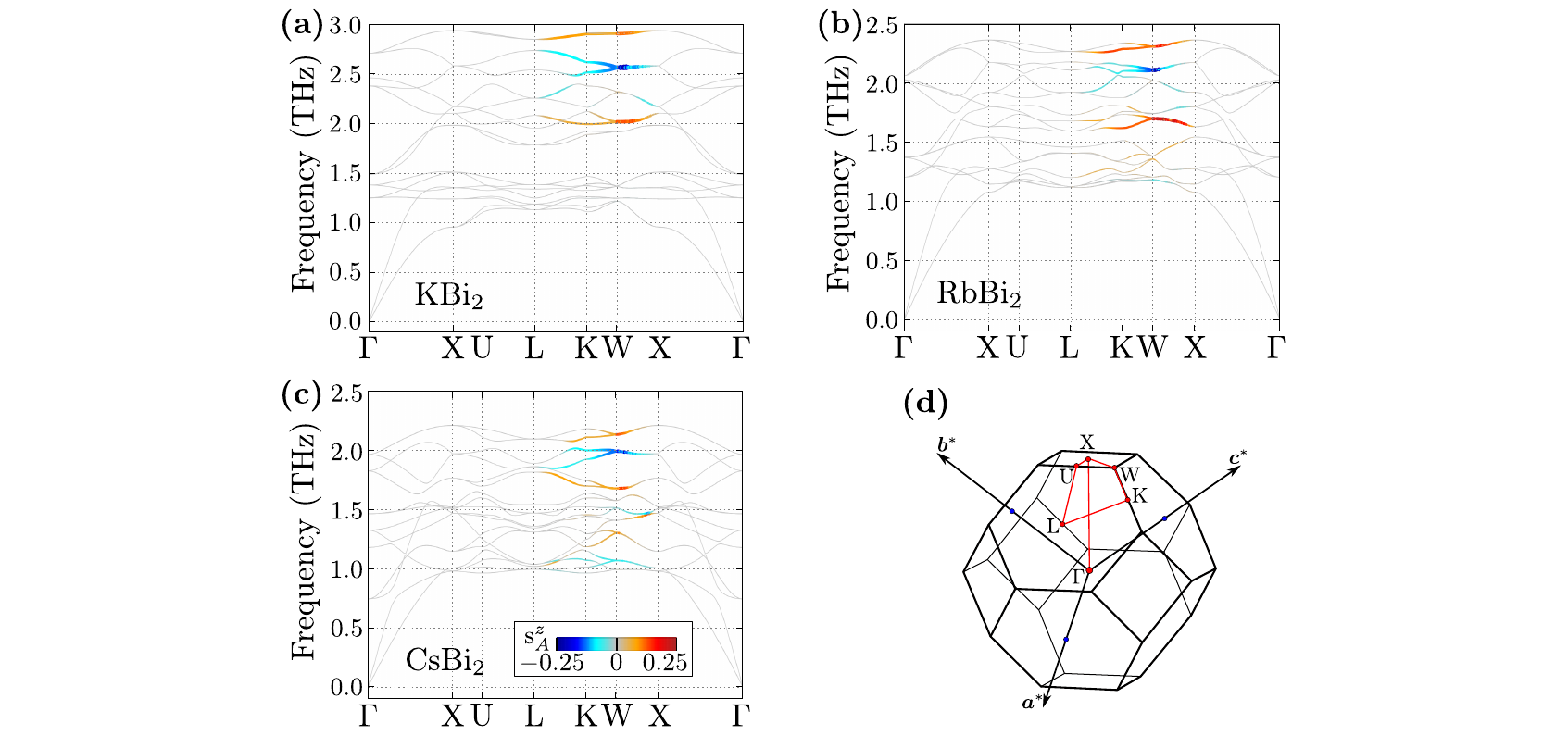}
\caption{
Phonon dispersion for the base systems $A$Bi$_{2}$ (as labeled), along high symmetry directions (a)--(c). 
Color code and linewidth correspond to the phonon circular polarization [for vibration in the plane perpendicular to the (111) direction] for one from the alkali atoms.
Panel (c) presents the discussed Brillouin zone and high symmetry points.}
\label{fig.clean}
\end{figure*}


\paragraph*{Motivation.---}
Chiral phonons can be characterized by the phonon circular polarization (PCP), or equivalently by the pseudo-angular momentum (PAM)~\cite{zhang.niu.15}.
Non-zero value of PCP (PAM) give information about the presence of the chiral phonons in the system [detailed discussion can be found in Supplementary Material (SM)~\footnote{See Supplementary Material at [URL will be inserted by publisher] for the computational details and the additional numerical results.}].
However, from experimental point of view, most interesting are materials, which in natural way posses a non-zero total PCP (PAM).
In such cases, the strain gradient~\cite{zhang.niu.15} or the temperature gradient~\cite{park.yang.20} can be source of phonon Hall effect.
The main condition to realize such systems is a broken inversion symmetry or/and time reversal symmetry~\cite{coh.19}.

In this Letter we discuss realization of the chiral phonon in the $A$Bi$_{2}$ ($A=$K, Rb, Cs) compounds~\cite{geert.dorn.61,emmerling.langin.04}, crystallized in the Laves phase (Fig.~\ref{fig.crys}).
Clean system, due to the presence of both inversion and time reversal symmetry, hosts chiral phonons with vanishing total PCP (PAM). 
We will show that the doped system can host chiral phonons with a non-zero total PCP (PAM), what that opens a new way to chiral phonon engineering and their experimental exploration.

In our study we used the {\it ab initio} techniques to study the dynamical properties of the (based and modified) systems.
During the study, we theoretically calculated and analyzed the phonon dispersions.
Firstly, this allows us to discuss the system stability in context of lattice dynamics.
Next, study of the phonon polarization vectors allow us to explore the behavior of chiral phonons, through the calculation of PCP (detailed can be found in the SM~\cite{Note1}).



\paragraph*{Calculations details.---}
The first-principle calculations were performed within density-functional theory (DFT) within the projector augmented-wave (PAW) method~\cite{blochl.94} implemented in the Vienna 
\textit{Ab initio} Simulation Package ({\sc vasp})~\cite{kresse.hafner.94,kresse.furthmuller.96,kresse.joubert.99}.
The exchange-correlation potential was obtained by the generalized 
gradient approximation (GGA) in the form proposed by Perdew, Burke, and Enzerhof (PBE)~\cite{pardew.burke.96}. 
We also investigated the impact of the spin--orbit coupling (SOC)~\cite{steiner.khmelevskyi.16} on the electronic structure.
The energy cut-off for the plane-wave expansion was equal to $350$~eV.
The optimization of the conventional cell was performed using a $10 \times 10 \times 10$ Monkhorst-Pack {\bf k}-grid~\cite{monkhorst.pack.76}. 
The structures were relaxed using the conjugate gradient technique with the energy convergence criteria set at $10^{-8}$~eV and $10^{-6}$~eV for the electronic and ionic iterations, respectively (the optimized lattice constant are in excellent agreement with the experimental values~\cite{emmerling.langin.04}, and are collected in Tab.~\ref{tab.lattice} in the SM~\cite{Note1}).
Symmetry of the structures were analyzed with {\sc FindSym}~\cite{stokes.hatch.05} and {\sc Seek-path}~\cite{hinuma.pizzi.17,togo.tanaka.18} packages.

The dynamical properties were calculated using the direct {\it Parlinski--Li--Kawazoe} method~\cite{parlinski.li.97}. 
Under this calculation, the interatomic force constants (IFC) are found from the forces acting on atoms when an individual atom is displaced.
The forces were obtained by the first-principle calculations with {\sc vasp} using conventional cell, and reduced $4 \times 4 \times 4$ {\bf k}-grid.
The phonon dispersion and polarization vectors analyses were performed using the {\sc Alamode} software~\cite{tadano.gohda.14,tadano.tsuneyuki.18}.
The mode symmetries at the $\Gamma$ point were found by the {\sc Phonopy} software~\cite{phonopy}.
Finally, the chiral phonons were studies by calculation of the PCP (for more details, see Sec.~\ref{app.pcp} in the SM~\cite{Note1}).


\paragraph*{Systems characterization.---}
Base compounds $A$Bi$_{2}$ ($A=$K, Rb, Cs) crystallize in the cubic Laves phase (space group Fd$\bar{3}$m, No.~$227$), with a pyrochlore lattice~\cite{geert.dorn.61,emmerling.langin.04} (left panel on Fig.~\ref{fig.crys}).
Alkali metals $A$ atom and Bi atom are in high symmetric Wyckoff position $8b$ ($3/8$,$3/8$,$3/8$) and $16c$ ($0$,$0$,$0$), respectively. 
The Bi atoms form corner sharing tetrahedrons, which are intertwined with the diamond structures formed by the Rb atoms.
Alternatively, the structure an be described as stacking of 3D kagome nets of Bi atoms.
The Rb atoms are located above the four hexagon faces of the capped tetrahedron of the 12 Bi atoms (right panel on Fig.~\ref{fig.crys}).
$A$Bi$_{2}$, similar to the related Bi compounds ($A$Bi or $A$Bi$_{3}$~\cite{sambongi.71,kushwaha.krizan.14,gornicka.gutowska.20,kinjo.kajino.16}), exhibit the superconducting properties~\cite{sun.liu.16,winiarski.wiendlocha.16,golab.wiendlocha.19,li.ikeda.22,gutowska.wiendlocha.22}.

In our study, we are focused on KRbBi$_{4}$ and RbCsBi$_{4}$, which can be realized when the one alkali atom (in primitive unit cell of the base $A$Bi$_{2}$) is replaced by one of another type.
Group analyses of these systems, after the DFT optimization procedure, show that both compounds have F$\bar{4}3$m symmetry (space group No.~$216$).
In this case, the alkali atoms are located at (non-equivalent) high symmetry Wyckoff position $4a$ ($0$,$0$,$0$) and $4c$ ($1/4$,$1/4$,$1/4$), while Bi atom in  position $16c$ ($x_\text{Bi}$,$x_\text{Bi}$,$x_\text{Bi}$), where $x_\text{Bi}$ is a free parameter (we found x$_\text{Bi}$ equal to approximately $\sim 3/8$ for both compounds).


\paragraph*{Phonon spectrum and flat band.---}
Let us start by giving a short description of the phonon spectrum.
All of the described compounds are stable, i.e. the phonon spectrum do not exhibit imaginary frequencies.
Exact analyses of the partial phonon density of states (Fig.~\ref{fig.dos} in the SM~\cite{Note1}) clearly show a decrease in the frequencies of vibrations related to the alkali metal, during substitution K$\rightarrow$Rb$\rightarrow$Cs, due to the increasing masses of the atoms.
Similar property is observed for the breathing mode of the Bi atoms tetrahedra.
This non-degenerate mode (with symmetry $A_\text{2u}$ for Fd$\bar{3}$m phase, and symmetry $A_\text{1}$ for F$\bar{4}3$m phase) shows an unusual decrease in frequency from $\sim 2.5$~THz for KBi$_{2}$ to $\sim 1.2$~THz for CsBi$_{2}$ (see Tab.~\ref{tab.irr} in the SM~\cite{Note1}). 
At the same time, oscillations of the heavy Bi atoms are mostly realized in the intermediate range of frequencies (approximately, between $1$~THz and $1.75$~THz).

For the compounds with ``light'' alkali atoms (KBi$_{2}$ and KRbBi$_{4}$), the separation of the modes in the frequency domain is observed -- the modes related to the alkali atoms are mostly observed at the higher frequencies.
Contrary to this, in the compounds containing the ''heavy'' alkali atoms (RbCsBi$_{4}$ and CsBi$_{2}$), the vibrational modes are strongly mixed between alkali and Bi atoms.
This type of the separation allow realized the phonon flat bands for KBi$_{2}$, at frequencies around $1.4$~THz [Fig.~\ref{fig.clean}(a)].
Formation of the phonon flat band is allowed due to the realization of the 3D kagome lattice of Bi atoms.
Indeed, the flat bands correspond to the strong peaks in the phonon density of states in the range of frequencies related to the Bi atoms [Fig.~\ref{fig.dos}(a) in the SM~\cite{Note1}].


\begin{figure}[!t]
\includegraphics[width=0.8\linewidth]{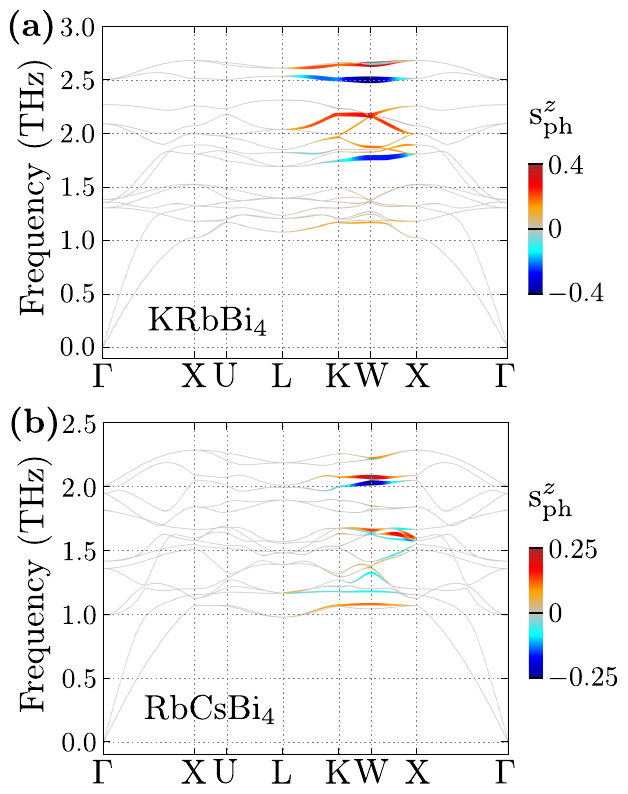}
\caption{
Phonon dispersion for the modified systems (as labeled), along the high symmetry directions [with the same marking as at Fig.~\ref{fig.doped}(d)]. 
Color code and linewidth correspond to the total phonon circular polarization [for vibration in the plane perpendicular to the (111) direction].
}
\label{fig.doped}
\end{figure}

\paragraph*{Chiral phonons and their properties.---}
Substitution of external atoms decrease the symmetry of the system but preserves the two- and three-fold rotational symmetry.
From this, the circular vibration of the atoms in both structures can be expected. 
Indeed, the PCP calculations (Fig.~\ref{fig.clean} and Fig.~\ref{fig.doped}) directly show the existence of chiral phonons.

In the case of Fd$\bar{3}$m phase, the chiral modes are realized by (opposite) circular vibrations of alkali atoms (located in two non-equivalent positions, let say $\mathcal{A}$ and $\mathcal{B}$, in primitive unit cell).
When atom in position $\mathcal{A}$ has positive circulation, the atom in position $\mathcal{B}$ has negative circulation, and {\it vice versa}.
Due to the equal masses of both the atoms, chiral modes are realized along the circles with the same radius in both positions.
However, due to the the inversion symmetry and time reversal symmetry, the chiral modes have zero total PCP.
In a natural way, the inversion symmetry can be broken by the introduction of external atoms in the base system $A$Bi$_{2}$, and that should lead to chiral phonons with non-zero PCP (or equivalently PAM)~\cite{coh.19}. This phenomena can be realized in
KRbBi$_{4}$ and RbCsBi$_{4}$, which can be made from KBi$_{2}$, RbBi$_{2}$, or CsBi$_{2}$, by the substitution of other alkali atoms.
Substitution of this type leads to a situation when diamond-like sublattice of the alkali atoms contains two different atoms (with different masses).
Then, contrary to the base compounds, the non-vanishing total PCP can be observed for several branches.
The related phonon dispersion are presented on Fig.~\ref{fig.doped}.

Here, we should also comment briefly on the chiral modes in the 3D kagome sublattice of Bi atoms.
In the 2D kagome lattice, the chiral modes can be realized by the breaking of spatial inversion symmetry~\cite{chen.wu.19}.
Thus, in the base $A$Bi$_{2}$ system the chiral modes are not realized in the Bi sublattice.
However, breaking the inversion symmetry in the modified compounds KRbBi$_{4}$ and RbCsBi$_{4}$ affect also the Bi sublattice (see Fig.~\ref{fig.sq} in the SM~\cite{Note1}).
Contrary to the vibrations of the alkali atoms, the Bi atoms within the chiral modes circulate along a ellipse (with small semi-minor axis).

In case of the modified systems, the non-zero PCP (PAM) arise as a consequence of the difference in the circular motion of the atoms (mostly, in two non-equivalent positions of the alkali atoms).
However, main properties of the circular motion of the alkali atoms in the system is preserved.
When the atom in position $\mathcal{A}$ has positive circulation and realize the motion along circle with some radius, the atoms in position $\mathcal{B}$ has negative circulation, and {\it vice versa}.

\begin{figure*}
\includegraphics[width=\linewidth]{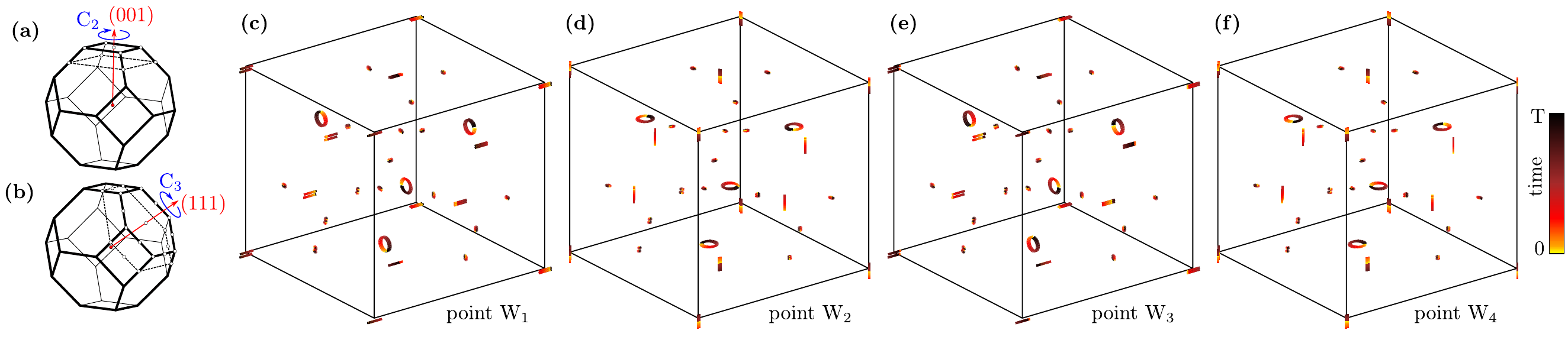}
\caption{
C$_{2}$ (a) and C$_{3}$ (b) rotation in the reciprocal space.
Visualization of the chiral mode in the conventional cell for W type wavevectros (c)-(f).
Color of line correspond to the time, during one period $T$ of the oscillation.
Results for 9th phonon mode realized by the Rb atom in KRbCs$_{4}$.
}
\label{fig.motion}
\end{figure*}

In Fig.~\ref{fig.motion} we present examples of the chiral modes realized for two non-equivalent W points.
Both wavectors are associated with the same square face of the Brillouin zone and posses 2-fold rotational symmetry [Fig.~\ref{fig.motion}(a)].
Due to the C$_{2}$ symmetry, the direction of circulation at one W point is a mirror reflection of its pair -- applying the C$_{2}$ symmetry leads to a change in the direction of circulation [cf. W$_{1}$ and W$_{3}$ presented in Fig.~\ref{fig.motion}(c) and Fig.~\ref{fig.motion}(e), or W$_{2}$ and W$_{4}$ presented on Fig.~\ref{fig.motion}(d) and Fig.~\ref{fig.motion}(f)].
This situation is similar to the simple 2D system with hexagonal symmetry~\cite{zhang.niu.15}, where the change of wavevector (from K to K') leads to the inversion of circulation.
However, contrary to the 2D hexagonal system, the chiral phonons do not vanish at the edge of the square face of the Brillouin zone (see Fig.~\ref{fig.sq} in the SM~\cite{Note1}).
This is associated with the additional properties of the system not observed in the 2D hexagonal system, i.e., change of the wavevecotr along the edge of the the square face of the Brillouin zone, lead also to the rotation of the plane within the chiral modes are realized.
This is well visible, when we compare the panels from (c) to (f) in Fig.~\ref{fig.motion}.
Similar behavior can be observed in case of the non-equivalent L points which are present in the hexagonal face of the the Brillouin zone and posses a three-fold rotational symmetry [Fig.~\ref{fig.motion}(b)].

The described behavior of the non-vanishing chiral modes at the edge of the Brillouin zone can be explained in a simple way.
In the 2D hexagonal system, change of phase of the phonon wavefunction must be realized along paths connecting $\Gamma$ points of neighboring  Brillouin zones (e.g. $\Gamma$--K--M--K'--$\Gamma$, where at the M point the chiral phonons are not realized).
This property is intrinsic, and in our case realized along paths $\Gamma$--X--$\Gamma$, or $\Gamma$--L--$\Gamma$, while X and L points are located exactly at the rotational symmetry axis.
As a consequence, at these points the chiral modes are not realized.
Indeed, in our case, on the square and hexagonal faces of Brillouin zone of the discussed systems, the chiral phonons are not realized only at the central X and L points.

Discussed changes of the phonon wavefunction phase (e.g. at the edge of the square face of the Brillouin zone, along W$_{1}$--W$_{2}$--W$_{3}$--W$_{4}$ path), is related only to the mentioned modification of the plane in which the chiral phonon modes are realized. 
At the same time, magnitudes of the circulations do not changed value -- this can be observed on the PCP plot (see Fig.~\ref{fig.sq} in the SM~\cite{Note1}).
The PCP changes value when we change one W point to another one,  (depending on the plane of the projection).


\paragraph*{Raman active modes.---}
In both type of symmetries, the modes at the $\Gamma$ point can be decomposed to seven irreducible representations: one non-degenerate $A$-type mode, one double-degenerate $E$-type mode, and five thyple-degeneated $T$-type modes (see Sec.~\ref{app.gamma} in the SM~\cite{Note1}).
Three of these modes are the Raman active (see Sec.~\ref{app.raman} in the SM~\cite{Note1}).
Using the Raman selective rules~\cite{loudon.01}, it is possible to distinguish the active modes, using the different backscattering configurations.
The same holds for both linear polarized (parallel and perpendicular configuration) and circular polarized (co-circular and cross-circular configuration) Raman spectra measurements.

Typically, in case of systems with hexagonal symmetry, the circular polarized Raman spectroscopy can be used to study the chiral modes~\cite{yin.ulman.21,du.tang.19}.
However, the method is limited only to the wavectors in vicinity of the $\Gamma$ point.
Our analyses of the realized chiral phonons (within the PCP calculations -- presented on Fig.~\ref{fig.clean} and Fig.~\ref{fig.doped}) clearly show that the chiral modes in discused compounds are realized around the edges of the Brillouin zone (e.g. along the L--K--W--X path).
From another site, the role of the external magnetic filed in realization of the chiral modes is well known~\cite{juraschek.spaldin.19,schaack.76,juraschek.neuman.22}.
The chiral modes can emerge in the presence of the magnetic field at the $\Gamma$ point, as a consequence of the decoupled doubly degenerate (Raman active) $E$ type modes, into two non-degenerated chiral modes with opposite circulation.
Frequencies of the $E$ type modes are collected in Tab.~\ref{tab.irr} in the SM~\cite{Note1}.


\paragraph*{Summary and outlook.---}
We discuss the realization of the chiral phonon modes in the cubic system KRbBi$_{4}$ and RbCsBi$_{4}$, based on the Laves phase of $A$Bi$_{2}$ ($A=$K, Rb, Cs).
Using the lattice dynamics, we show that the two new compounds (KRbBi$_{4}$ and RbCsBi$_{4}$) can be stable, and crystallize with the F$\bar{4}3$m symmetry. 
The initial $A$Bi$_{2}$ systems poses both inversion symmetry and time reversal symmetry and as a consequence the chiral phonons with total zero phonon circular polarization (pseudo angular momentum) can be realized.
In this case the alhali atoms undergo circular motions with opposite circulations.
Contrary to this, in the proposed KRbBi$_{4}$ and RbCsBi$_{4}$ compounds, the inversion symmetry is broken, and the chiral phonon modes with non-zero phonon circular polarization (pseudo angular momentum) emerge.
Additionally, the chiral modes are realized for the wavevectors at the edge of the Brillouin zone, which posses two- and three-fold rotational symmetry.
The vanishing of the chiral modes is observed only for X and L points, which are located at the rotational symmetry axis.
Realiztion of chiral phonons with a non-zero PAM in the reported compunds opens up a new way to engineering materials for phonon Hall effect.

\begin{acknowledgments}
\paragraph*{Acknowledgments.---}
Some figures and movies in this work were rendered using 
{\sc Vesta}~\cite{momma.izumi.11} and {\sc VMD}~\cite{vmd} software.
We thank Sylwia Gutowska and Bartl\l{}omiej Wiendlocha for valuable comments and discussions.
S.B. is grateful to IT4Innovations (V\v{S}B-TU Ostrava) for hospitality during a part of the work on this project. 
This work was supported by the National Science Centre (NCN, Poland) under grants No. 
2017/25/B/ST3/02586 (S.B.) 
and
2021/43/B/ST3/02166 (A.P.). 
A.P. appreciates funding in the frame of scholarships of the Minister of Science and Higher Education (Poland) for outstanding young scientists (2019 edition, no. 818/STYP/14/2019).
\end{acknowledgments}

\bibliography{biblio.bib}

\newpage

\clearpage
\newpage

\onecolumngrid

\begin{center}
  \textbf{\Large Supplemental Material}\\[.2cm]
  \textbf{\large Chiral phonon in the cubic system based on the Laves phase of $A$Bi$_{2}$ ($A=$K, Rb, Cs)}\\[.2cm]
  Surajit Basak, Przemys\l{}aw Piekarz, and Andrzej Ptok\\[.2cm]
  {\itshape
  	\mbox{Institute of Nuclear Physics, Polish Academy of Sciences, W. E. Radzikowskiego 152, PL-31342 Krak\'{o}w, Poland}
  }
\\
(Dated: \today)
\\[1cm]
\end{center}

\setcounter{equation}{0}
\renewcommand{\theequation}{S\arabic{equation}}
\setcounter{figure}{0}
\renewcommand{\thefigure}{S\arabic{figure}}
\setcounter{section}{0}
\renewcommand{\thesection}{S\Roman{section}}
\setcounter{table}{0}
\renewcommand{\thetable}{S\arabic{table}}
\setcounter{page}{1}

\onecolumngrid

In this Supplemental Material we present additional results, in particular concerning:
\begin{itemize}
\item Lattice constant found from the system optimization during the DFT calculations (Tab.~\ref{tab.lattice}).
\item Transformation of the primitive unit cell to the conventional cell (Sec.~\ref{app.conv}).
\item Direct expression to calculate phonon cirucalr polarization (Sec.~\ref{app.pcp}).
\item Irreducible representations of modes at $\Gamma$ point (Sec.~\ref{app.gamma}) and their frequencies (Tab.~\ref{tab.irr})
\item Polarization selection rules for Raman scattering (Sec.~\ref{app.raman}).
\item Calculated total and partial phonon density of states  (Fig.~\ref{fig.dos}).
\item Phonon circular polarization at the edge of W--X plane (Fig.~\ref{fig.sq}).
\end{itemize}


\begin{table}[!h]
\caption{
\label{tab.lattice}
The lattice constants of the discussed compounds (as labeled).
}
\begin{ruledtabular}
\begin{tabular}{cccccc}
& KBi$_{2}$ & KRbBi$_{4}$ & RbBi$_{2}$ & RbCsBi$_{4}$ & CsBi$_{2}$ \\ 
\hline 
symmetry & Fd$\bar{3}$m & F$\bar{4}3$m & Fd$\bar{3}$m & F$\bar{4}3$m & Fd$\bar{3}$m \\
a (\AA) & $9.6642$ & $9.7177$ & $9.7702$ & $9.8421$ & $9.9112$
\end{tabular}
\end{ruledtabular}
\end{table}


\section{Construction of conventional cell}
\label{app.conv}

Primitive unit cell (with lattice vectors ${\bm a}_{p}$,
${\bm b}_{p}$, and ${\bm c}_{p}$) can be transform to conventional one (described by vectors ${\bm a}_{c}$, ${\bm b}_{c}$, and ${\bm c}_{c}$), by transformation:
\begin{eqnarray}
\left( \begin{array}{c}
{\bm a}_{c} \\ 
{\bm b}_{c} \\ 
{\bm c}_{c}
\end{array} \right) = 
\left( \begin{array}{rcc}
1 & -1 & -1 \\ 
1 & -1 & 1 \\ 
-1 & -1 & 1
\end{array}  \right) 
\left( \begin{array}{c}
{\bm a}_{p} \\ 
{\bm b}_{p} \\ 
{\bm c}_{p}
\end{array} \right) .
\end{eqnarray}


\section{Phonon circular polarization}
\label{app.pcp}

The lattice dynamics of the system is based on the study of the dynamical matrix:
\begin{eqnarray}
D_{\alpha\beta}^{jj'} \left( {\bm q} \right) \equiv \frac{1}{\sqrt{m_{j}m_{j'}}} \sum_{n} \Phi_{\alpha\beta} \left( j0, j'n \right) \exp \left( i {\bm q} \cdot {\bm R}_{j'n} \right) ,
\label{eq.dyn_mat}
\end{eqnarray}
where ${\bm q}$ is the phonon wave vector and $m_{j}$ denotes the mass of $j$-th atom.
Here, $\Phi_{\alpha\beta} \left( j0, j'n \right)$ is the interatomic force constant tensor ($\alpha$ and $\beta$ denotes the direction index, i.e., $x$, $y$, and $z$) between $j$-th and $j'$-th atoms located in the initial (0) and $n$-th primitive unit cell. 
The phonon spectrum for a given wave vector ${\bm q}$ is specified by the eigenproblem of the dynamical matrix~(\ref{eq.dyn_mat}), i.e.,
\begin{eqnarray}
\omega^{2}_{\varepsilon{\bm q}} \evec_{\varepsilon{\bm q}\alpha j} = \sum_{j'\beta} D_{\alpha\beta}^{jj'} \left( {\bm q} \right) \evec_{\varepsilon{\bm q}\beta j'} .
\end{eqnarray}
Here, the $\varepsilon$ branch describes the phonon with $\omega_{\varepsilon{\bm q}}$ frequency and polarization vector $\evec_{\varepsilon{\bm q}\alpha j}$.
Each $\alpha j$ component of the polarization vector denotes displacement of $j$-th atom in $\alpha$-th direction.

To study of the chiral phonons, we performed theoretical analysis of the polarization vectors, analogically to the previous study~\cite{zhang.niu.15}.
We can introduce a new basis defined as:
$|R_{1}\rangle \equiv \frac{1}{\sqrt{2}} \left( 1 \; i \; 0 \cdots 0 \right)^\text{T}$, 
$|L_{1}\rangle \equiv \frac{1}{\sqrt{2}} \left( 1 \; -i \; 0 \cdots 0 \right)^\text{T}$, 
$|Z_{1}\rangle \equiv \left( 0 \; 0 \; 1 \cdots 0 \right)^\text{T}$; $\ldots$; 
$|R_{j}\rangle \equiv \frac{1}{\sqrt{2}} \left( 0 \cdots 1 \; i \; 0 \cdots 0 \right)^\text{T}$, $|L_{j}\rangle \equiv \frac{1}{\sqrt{2}} \left( 0 \cdots 1 \; -i \; 0 \cdots 0 \right)^\text{T}$, $|Z_{j}\rangle \equiv \left( 0 \cdots 0 \; 0 \; 1 \cdots 0 \right)^\text{T}$; $\ldots$; i.e., 
components corresponding to the $i$-th atom are replaced by the circular polarization vectors (i.e. $\sigma^{\pm} = ( 1 \; \pm i \; 0)$ for circulation in $xy$ plane -- however, form $\sigma^{\pm}$ can be chose  arbitrary).
The vector $| R_{i} \rangle$ and $| L_{i} \rangle$ correspond to the right-handed polarization (RHP) and the left-handed polarization (LHP), respectively.

In this basis, each polarization vector $\evec \equiv \evec_{\varepsilon{\bm q}\alpha j}$, can be represented as:
\begin{eqnarray}
\evec = \sum_{j} \left( \alpha^{R}_{j} | 
R_{j} \rangle + \alpha^{L}_{j} | L_{j} \rangle + \alpha^{Z}_{j} | Z_{j} \rangle \right) ,
\end{eqnarray}
where $\alpha^{V}_{j} = \langle V_{j} | \evec \rangle$, for $V \in \{ R, L, Z \}$ and \mbox{$j \in \{ 1 , 2, \cdots, N \}$} ($N$ is a total number of atoms in a primitive unit cell).
Additionally, we can define the phonon circular polarization operator $\hat{S}^{z}_{ph}$ along the $z$ direction as:
\begin{eqnarray}
\hat{S}^{z}_{ph} \equiv \sum_{j} \hat{S}_{j}^{z} = 
\sum_{j} \left( | {R}_{j} \rangle \langle R_{j} | - | L_{j} \rangle \langle L_{j} | \right) , 
\end{eqnarray}
where $\hat{S}_{i}^{z}$ is the phonon circular polarization operator at site $i$.
From this, the phonon circular polarization $s_{j}^{z}$ of $j$-th atom can be expressed as
\begin{eqnarray}
s_{j}^{z} = \hslash \evec^{\dagger} \hat{S}_{j}^{z} \evec = 
\hslash \left( | \alpha^{R}_{j} |^{2} - | \alpha^{L}_{j} |^{2} \right) .
\end{eqnarray}
It corresponds to the phonon angular momentum along the $z$ direction~\cite{zhang.niu.15}. 
However, in the general case, we can discuss the angular momenta along some arbitrary direction. 
For $s_{j}^{z} > 0$ ($s_{j}^{z} < 0$) the phonon mode has RHP (LHP), while for $s_{j}^{z} = 0$ the phonon mode is linearly polarized. 
Finally, the total phonon circular polarization $s_{ph}^{z}=\sum_j s_j^z$ denotes polarization of a whole system.


\section{The modes analyses at $\Gamma$ point}
\label{app.gamma}

The modes at the $\Gamma$ point can be decomposed into the irreducible representations as follows:
\begin{itemize}
\item for the case of symmetry Fd$\bar{3}$m (SG:227):
\begin{eqnarray}
\Gamma_\text{acoustic} = T_\text{1u} , \quad \text{and} \quad \Gamma_\text{optic} = A_\text{2u} + E_\text{u} + T_\text{2u} + T_\text{2g} + 2 T_\text{1u} .
\end{eqnarray}
\item for the case of symmetry F$\bar{4}3$m (SG:216):
\begin{eqnarray}
\Gamma_\text{acoustic} = T_\text{2} , \quad \text{and} \quad \Gamma_\text{optic} = A_\text{1} + E + T_\text{1} + 3 T_\text{2} .
\end{eqnarray}
\end{itemize}

\begin{table}[!h]
\caption{
\label{tab.irr}
Irreducible representations of modes at $\Gamma$ point.
}
\begin{ruledtabular}
\begin{tabular}{cl|cl|cl|cl|cl}
\multicolumn{2}{c}{KBi$_{2}$} & \multicolumn{2}{c}{KRbBi$_{4}$} & \multicolumn{2}{c}{RbBi$_{2}$} & \multicolumn{2}{c}{RbCsBi$_{4}$} & \multicolumn{2}{c}{CsBi$_{2}$} \\
Freq. (THz) & Rep. & Freq. (THz) & Rep. & Freq. (THz) & Rep. & Freq. (THz) & Rep. & Freq. (THz) & Rep. \\
\hline 
$1.254$ & $T_\text{2u}$ & $1.310$ & $T_\text{1}$ & $1.203$ & $T_\text{1u}$ & $0.991$ & $T_\text{2}$ & $0.745$ & $T_\text{1u}$ \\
$1.384$ & $E_\text{u}$ & $1.357$ & $T_\text{2}$ & $1.368$ & $E_\text{u}$ & $1.357$ & $E$ & $1.179$ & $A_\text{2u}$ \\
$1.489$ & $T_\text{1u}$ & $1.387$ & $E$ & $1.370$ & $T_\text{2u}$ & $1.417$ & $T_\text{1}$ & $1.325$ & $E_\text{u}$ \\
$2.379$ & $T_\text{1u}$ & $2.094$ & $T_\text{2}$ & $2.020$ & $T_\text{1u}$ & $1.812$ & $A_\text{1}$ & $1.463$ & $T_\text{2u}$ \\
$2.461$ & $A_\text{2u}$ & $2.269$ & $A_\text{1}$ & $2.023$ & $A_\text{2u}$ & $1.945$ & $T_\text{2}$ & $1.824$ & $T_\text{2g}$ \\
$2.707$ & $T_\text{2g}$ & $2.502$ & $T_\text{2}$ & $2.057$ & $T_\text{2g}$ & $2.005$ & $T_{2}$ & $1.962$ & $T_\text{1u}$
\end{tabular}
\end{ruledtabular}
\end{table}


\begin{figure*}[!t]
\includegraphics[width=0.6\linewidth]{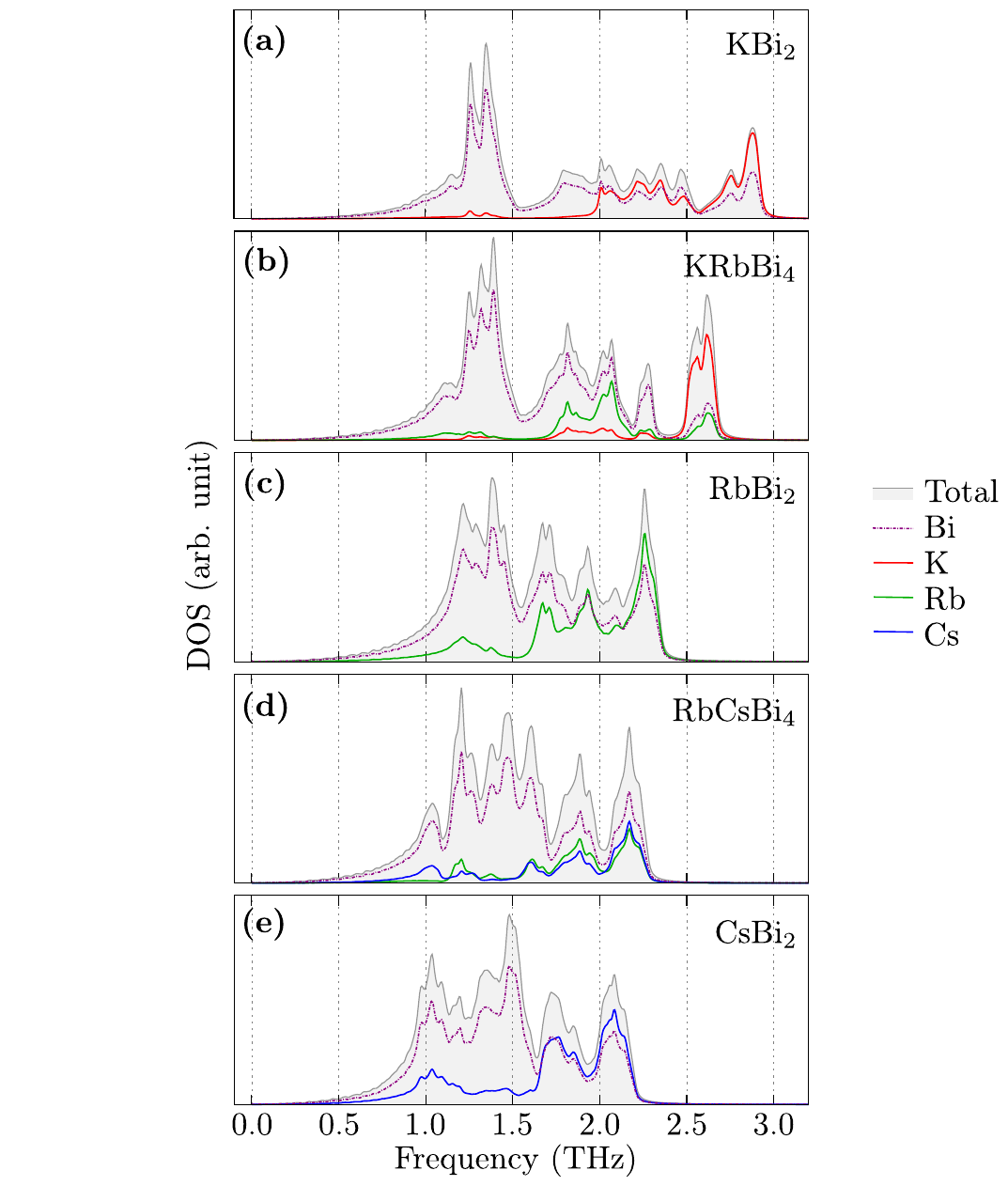}
\caption{
Total and partial density of states (DOS) for discuss systems (as labeled).
}
\label{fig.dos}
\end{figure*}


\section{Polarization selection rules for Raman scattering}
\label{app.raman}

According to the group theory, the Raman tensor for the Fd$\bar{3}$m space group takes the following forms:
\begin{eqnarray}
\label{eq.raman} &R \left( A_\text{1g} \right) =  \left( \begin{array}{ccc}
a & 0 & 0 \\ 
0 & a & 0 \\ 
0 & 0 & a
\end{array} \right) ; & R \left( E_\text{g}^\text{I} \right) = \left( \begin{array}{ccc}
b & 0 & 0 \\ 
0 & b & 0 \\ 
0 & 0 & -2b
\end{array} \right); \; R \left( E_\text{g}^\text{II} \right) =
\left( \begin{array}{ccc}
-\sqrt{3} b & 0 & 0 \\ 
0 & \sqrt{3} & 0 \\ 
0 & 0 & 0
\end{array} \right) ; \\
\nonumber &R \left( T_\text{2g}^\text{I} \right) =  \left( \begin{array}{ccc}
0 & 0 & 0 \\ 
0 & 0 & d \\ 
0 & d & 0
\end{array} \right) ; & R \left( T_\text{2g}^\text{II} \right) = \left( \begin{array}{ccc}
0 & 0 & d \\ 
0 & 0 & 0 \\ 
d & 0 & 0
\end{array} \right); \; R \left( T_\text{2g}^\text{III} \right) =
\left( \begin{array}{ccc}
0 & d & 0 \\ 
d & 0 & 0 \\ 
0 & 0 & 0
\end{array} \right) .
\end{eqnarray}
Similarly, the Raman tensors take the same form (for $A_\text{1}$, $E$ and $T_\text{2}$ modes) for the Raman active modes in the system with F$\bar{4}3$m symmetry, .

The non-resonant Raman scattering intensity depends in general on the directions of the incident and scattered light relative to the principal axed of the crystal. 
It is expressed by Raman tensor $R$, relevant for a given crystal symmetry, as~\cite{loudon.01}:
\begin{eqnarray}
\label{eq.intens} I \propto | e_{i} \cdot R \cdot e_{s} |^{2} ,
\end{eqnarray}
where $e_{i}$ and $e_{s}$ are the polarization vectors of the incident and scattered light, respectively.
The Raman tensors $R$ are presented in (\ref{eq.raman}).

In the backscattering configuration, $e_{i}$ and $e_{s}$ are placed within the $xy$ plane.
The polarization vectors for linearly polarized light in the $x$ and $y$ directions are \mbox{$e_{x} = \left( 1 \; 0 \; 0 \right)$} and \mbox{$e_{y} = \left( 0 \; 1 \; 0 \right)$}, respectively.
Similarly, the polarization vectors for left $\sigma^{+}$ and right $\sigma^{-}$ circularly polarized light are \mbox{$\sigma^{\pm} = \frac{1}{\sqrt{2}} \left( 1 \; \pm i \; 0 \right)$}.
Using Eq.~(\ref{eq.intens}) and the Raman tensors~(\ref{eq.raman}), we can determine the selection rules and Raman intensities for various scattering geometries.
Tab.~\ref{tab.ramanselect} summarizes the Raman response in the backscattering geometry for four polarization configurations.

\begin{table}[!t]
\caption{
\label{tab.ramanselect}
Selection rules for Raman-active modes.
}
\begin{ruledtabular}
\begin{tabular}{ccccc}
configuration & $e_{x}$ in $e_{x}$ out & $e_{x}$ in $e_{y}$ out & $\sigma^{+}$ in $\sigma^{+}$ out & $\sigma^{+}$ in $\sigma^{-}$ \\
 & (linear $\parallel$) & (linear $\perp$) & (cocircular) & (cross-circular) \\
\hline 
$A_\text{1g}$ / $A_{1}$ & $a^{2}$ & $0$ & $a^{2}/2$ & $0$ \\
$E_\text{g}$ / $E$ & $b^{2}$ & $0$ & $b^{2}/2$ & $3b^{2}/2$ \\
$T_\text{2g}$ / $T_\text{2}$ & $0$ & $0$ & $0$ & $2d^{2}$
\end{tabular}
\end{ruledtabular}
\end{table}


\begin{figure}[!h]
\includegraphics[width=\linewidth]{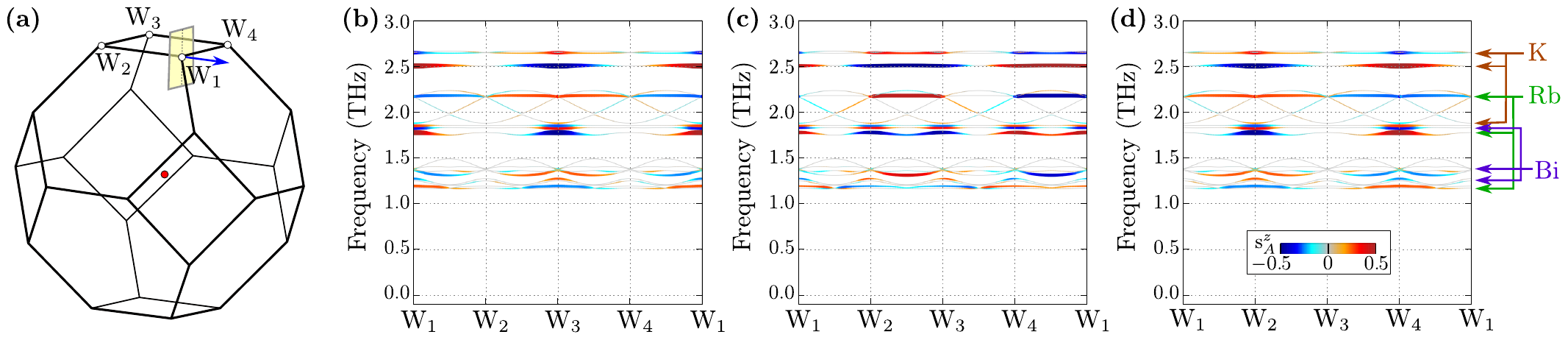}
\caption{
Position of the discussed wavevectors (W type points) in the reciprocal space.
Phonon circular polarization for the circulation realized in the different planes: perpendicular to $\hat{x}$ (b) [yellow plane shown on (a)], perpendicular to $\hat{x}+\hat{y}$ (c), and perpendicular $\hat{y}$ (d).
Information about orientated frequencies related to the different atomic vibrations are presented from right side of panel (d).
}
\label{fig.sq}
\end{figure}

\end{document}